\documentclass[aps,prl,twocolumn,nofootinbib]{revtex4}
\usepackage{graphics,graphicx,epsfig}
\usepackage{amsfonts,amsmath,amssymb}
\def\tphi{\mathcal T_\phi }

\def\b2hat{ {\hat b}_2 }

\begin{document}

\title{The Angular Tension of Black Holes}
\author{David Kastor}
\author{Jennie Traschen}
\affiliation{Department of Physics, University of Massachusetts, Amherst, MA 01003}

\begin{abstract} 
Angular tension is an ADM charge that contributes a work term to the first law of black hole mechanics when the range of an angular coordinate is varied and leads to a new Smarr formula for stationary black holes.  
A phase diagram for singly-spinning $D=5$ black holes shows that angular tension resolves the degeneracies between spherical black holes and (dipole) black rings and captures the physics of the black ring balance condition.
Angular tension depends on the behavior of the metric at rotational axes and we speculate on its relation to rod/domain structure characterizations of higher dimensional black holes and black hole uniqueness theorems.
\end{abstract}


\maketitle

\noindent\textbf{Introduction:} In this paper we identify a new property of black holes called angular tension that provides a useful tool for mapping the complicated spaces of higher dimensional black hole solutions.  Angular tension  is  closely related to the well known linear tension \cite{Traschen:2001pb,Townsend:2001rg,Harmark:2004ch} of Kaluza-Klein black holes, which 
helps characterize black holes in spacetimes with  compact dimensions \cite{Harmark:2003dg}.  Both quantities enter the first law of black hole mechanics through work terms.  Linear tension gives the variation in the black hole mass due to a change in the length of a compact dimension \cite{Townsend:2001rg,Harmark:2003dg,Kastor:2006ti}, while angular tension determines the change in mass when the range of an angular coordinate is varied.  Moreover both quantities are ADM charges \cite{Arnowitt:1962hi} of the black hole, defined via the same construction used to obtain the black hole mass and angular momentum. 

While stationary, vacuum black holes in $D=4$ are uniquely characterized 
by their mass and angular momentum, the spaces of higher dimensional solutions have  much richer structures (see \textit{e.g.} \cite{Emparan:2008eg,Ida:2011jm,Chrusciel:2012jk,Hollands:2012xy}). First there are more rotational degrees of freedom, with a total of  $[(D-1)/2]$ independent angular momenta, where $[x]$ denotes the integer part of $x$ and all $D$ dimensions are assumed to be non-compact \cite{Myers:1986un}.    More radically the event horizon, which must have spherical topology in $D=4$ \cite{Hawking:1971vc}, is less constrained in higher dimensions.  In $D=5$ for example, black ring solutions with toroidal horizons \cite{Emparan:2001wn}
are known, as well as spherical black holes \cite{Myers:1986un}.  
Moreover, multiple solutions exist having  the same mass and angular momenta, and hence these quantities no longer uniquely characterize a black hole.  We will see that angular tension removes at least part of this degeneracy.

We further expect that angular tension will be useful in distinguishing solutions that have different amounts of rotational symmetry.
All explicitly known stationary black hole solutions are symmetric in each independent rotational plane.  In $D=4$ this symmetry follows from the black hole rigidity theorem  \cite{Hawking:1971vc}.  However, the theorem in higher dimensions  \cite{Hollands:2006rj} continues to guarantee only a single rotational symmetry
and  solutions with smaller amounts of symmetry are conjectured to exist \cite{Reall:2002bh}.  
In the Kaluza-Klein case, phase diagrams plotting linear tension against black hole mass \cite{Harmark:2003dg} have been used to map out the transitions between branches of solutions that are translationally invariant in the compact direction (uniform black string and Kaluza-Klein bubble) and those that break this symmetry (non-uniform black strings and localized black holes).  Angular tension should play a similar role for charting branches that are uniform and those that are non-uniform in the azimuthal directions.

One can begin by looking at the angular tension of static black holes. 
We single out one of the azimuthal coordinates designated as $\phi$ and allow its range to vary in proportion to a parameter $k$, so that $0\le\phi <2\pi k$.  Based on the explicit solutions (see {\it e.g.} reference \cite{Myers:1986un}) one can write an equation of state for the black hole that expresses its mass in terms of its horizon area $A$ and $k$
\begin{equation}
M = {(D-2)\over 16\pi}\, (\Omega_{D-2} k)^{1\over D-2}\, A^{D-3\over D-2}
\end{equation}
where $\Omega_n$ is the area of the unit $n$-sphere.  Defining the angular tension $\tphi$ to be the variation of the mass with respect to the range of the angular coordinate $\phi$, with the horizon area held fixed, we find that for a static black hole
\begin{equation}\label{static}
\tphi = {M\over (D-2) 2\pi k},
\end{equation}
This result is analogous to the linear tension of a uniform black string, which is given by $M/(D-2)L$ where $L$ is the length of the compact direction and $D$ is the total number of noncompact dimensions.  We next see that $\tphi$  arises as an ADM charge.


\vskip 0.1in\noindent\textbf{ADM charges:} The ADM prescription for defining a conserved charge in general relativity involves choosing a hypersurface $\Sigma$ and also a symmetry or Killing vector $\xi^a$ \cite{Abbott:1981ff}.  Different choices lead to different charges.  Starting with a background spacetime for which the symmetry is exact, one considers perturbations that may preserve the symmetry only in an asymptotic limit (see references \cite{Traschen:1984bp,Sudarsky:1992ty,Traschen:2001pb} for more detailed accounts).  
In this section we consider perturbations off Minkowski generated by a stress-energy source $T_{ab}$.
Using the Einstein constraint equations on $\Sigma$ one can construct a Gauss' law type relation $D_a B^a=16\pi\epsilon\, \xi^aT_a{}^b\, n_b$ for a certain vector field $B^a$ constructed from the metric perturbation and the Killing vector.   Here $n_a$ is the normal to $\Sigma$ and $\epsilon = n_an^a=\pm 1$.  The ADM charge $Q$ may then be written equivalently as a surface integral or a volume integral
\begin{equation}\label{ADM}
Q= -(1/16 \pi)\int _{\partial\Sigma} da_c B^c = -\epsilon \int_\Sigma dv\, \xi^a T_a{}^b\, n_b
\end{equation}
Working with a flat background spacetime, 
the ADM mass is obtained by taking a constant time hypersurface and the time translation Killing vector.  It is then equal to the volume integral of the energy density $T_{tt}$.
In a Kaluza-Klein spacetime with compact spatial coordinate $z$, the linear tension \cite{Traschen:2001pb,Townsend:2001rg,Harmark:2004ch} is similarly obtained by taking a hypersurface of constant $z$ and the background $z$-translation Killing vector, and is  equal to minus the integral of pressure in the $z$-direction $T_{zz}$.

Angular tension $\tphi$ is the ADM charge associated with a hypersurface of constant azimuthal coordinate $\phi$ together with the rotational Killing vector $(\partial/\partial\phi)^a$ and is equal to minus the integral of the pressure in the $\phi$-direction $T_{\phi\phi}$.  It is this minus sign that justifies the terminology `tension', rather than `pressure'.  With this choice of $\Sigma$ the expressions in (\ref{ADM})  include integration over the time direction.  
However, if we restrict our interest to stationary spacetimes, then this integral will be trivial and may be supressed.  The quantity $\tphi$ is then strictly speaking a tension per unit time.  

We are interested in the boundary integral expression for $\tphi$ that follows from (\ref{ADM}).  In addition to the usual boundary at infinity a constant $\phi$ hypersurface has a boundary at the rotational axis, which is really a $(D-3)$-dimensional plane.  There will generally be a nontrivial contribution to $\tphi$ from the integral at this  boundary.  The angular tension can then be written as $\tphi=\tphi^\infty+\tphi^{axis}$, where $\tphi^*=-(1/16 \pi)\int _{\partial\Sigma_*} da_c B^c$.
This may appear to be a drawback.  However, where uniqueness theorems for higher dimensional stationary black holes have been proved  \cite{Hollands:2007qf}, additional data on rotational axes is required to fully specify a solution.  It may be possible to recast this data in the framework of ADM charges using angular tension and its generalizations.

We start with the contribution from infinity.   To further simplify matters we assume that angular momenta in the other planes of rotation vanish and accordingly take the fall-off conditions on the metric  at large radius to be 
\begin{eqnarray}\label{falloff}
&& ds^2 \simeq   (-1 + {c_t \over  r^{D-3}} )dt^2  + 2 {c_{t\phi}  \over r^{D-3} }\, \sin^2 \theta dt d\phi \\ &&  +
    (1 + {c_r \over r^{D-3} }) dr^2+   r^2 (d\theta  ^2  + \sin^2 \theta d\phi  ^2 +\cos^2\theta\,  d\Omega_{D-4}^2)\nonumber  
 \end{eqnarray}
where $c_t$, $c_r$, $c_{t\phi}$ are constants.  The $\phi$ coordinate is assumed to have the range $0\le\phi <2\pi k$.
The ADM mass and angular momentum are given by 
$M= k\Omega_{D-2}(D-2)c_r/16\pi$ and $J=-k \Omega_{D-2}c_{t\phi}/8\pi$.
One finds that the contribution to the angular tension from the boundary at infinity is given by
\begin{equation}\label{infinity}
\tphi^\infty = - {\Omega_{D-2}\over 16\pi}\, \left( (D-2)c_t - (D-3)c_r\right)
\end{equation}
Since $c_t=c_r$ for asymptotically flat vacuum solutions, one has $\tphi^\infty=-M/(D-2)2\pi k$.
Note that this result is actually minus the angular tension of a static black hole as given in (\ref{static}).  The difference represents the axis contribution.

To compute $\tphi^{axis}$
we need to specify the behavior of the metric near this axis.  
This is determined by requiring that the metric $g_{ab}$ be regular at the axis when the parameter $k=1$.  A value $k\ne 1$ indicates the presence of a conical singularity in the rotational plane.  Including this singularity is the price to be paid for allowing the angular range to vary.  At the end we will set $k=1$.
The relevant terms in the metric near the axis at $\theta=0$  are then given by 
\begin{eqnarray}\label{nearaxis} 
ds^2 &&\simeq  g_{tt} (r,y^k)dt^2 + g_{rr}(r,y^k)dr^2 
\\  \nonumber
&&  +r^2(d\theta^2+k^2 \theta^2 d\phi^2+  \gamma_{ij}(r,y^k)dy^idy^j)
\end{eqnarray}
where $y^i$ and $\gamma_{ij}(r,y^k)$ are coordinates and metric on the not necessarily round  $(D-4)$-sphere. In this section all components  in (\ref{nearaxis}) are assumed to be perturbatively close to the flat metric.
Computing the boundary vector component $B^\theta$ in this near axis limit yields the result
\begin{equation}\label{axis}
\tphi^{axis}= -{1\over 8\pi}\int_{axis}drdy^kr^{D-4}\sqrt{\bar\gamma}\, h .
\end{equation}
Here $h$
is the trace of the metric perturbation and is proportional to the perturbation to the intrinsic volume element on the axis.   
The  contribution $\tphi^\infty$ in (\ref{infinity}) has  the usual form for an ADM charge, {\it i.e.} an expression in terms of falloff coefficients in the far field.  However, the axis contribution $\tphi^{axis}$ in (\ref{axis}) necessarily involves the metric in the interior of the spacetime.   

We can get insight into $\tphi^{axis}$ by considering  a spherical star.  Outside the star the metric is Schwarzchild and the quantity $h$ in (\ref{axis}) vanishes.  The quantity $\tphi^{axis}$ then receives contributions only from the portion of the axis interior to the star and can be shown to be proportional to the integrated Newtonian potential.  Gauss' law (\ref{ADM}) then says that this interior contribution plus $\tphi^\infty$ give the integral of the pressure $p_\phi$ of the the star.

\vskip 0.1in\noindent\textbf{Black hole and the first law:} 
The first law construction also makes use of the Gauss' law relation presented above, but now applied to a stationary black hole background, with a constant time hypersurface $\Sigma$ and the horizon generating Killing vector $\xi^a = (\partial/\partial t)^a+\Omega(\partial/\partial\phi)^a$ where $\Omega$ is the horizon angular velocity (see \cite{Traschen:1984bp,Sudarsky:1992ty,Traschen:2001pb} for detailed accounts).  Now we take $T_{ab}=0$, but the black hole horizon introduces an interior component of $\partial\Sigma$.
For $\delta k=0$, the Gauss' law relation yields the usual first law
$\delta M=\kappa\delta A/8\pi + \Omega\delta J$, with the term $\kappa\delta A$ coming from the horizon and the terms proportional to variations in the mass and angular momentum coming from infinity.

Now consider the new terms that arise when the parameter $k$ specifying the range of $\phi$  is varied.
This  perturbation will be singular on the azimuthal axis.  So as above we expect a contribution  coming from a boundary surrounding the axis, as well as contributions coming from infinity and the black hole horizon.  We can summarize the computation by writing $I_{h}+ I_{axis} + I_\infty = 0$,
where the $I_\star=-(1/16\pi)\int_{\partial\Sigma_\star} da_c B^c$ are integrals on the different components of the boundary of $\Sigma$.
The boundary integral on the horizon is straightforward, continuing to be given by $I_{h} =-\kappa \delta A/8\pi$
but $\delta A$ now includes a contribution proportional to $\delta k$.

The boundary integral at infinity can be written as $I_\infty = I_\infty |_{\delta k=0} +I^\prime_\infty$, where 
$I_\infty |_{\delta k=0} = (\delta M-\Omega\delta J)|_{\delta k=0}$ and $I^\prime_\infty$ is proportional to $\delta k$.  
Because $M$ and $J$ are given by integrals over the sphere, they are each proportional to $k$, 
so that $\delta M= \delta M|_{\delta k=0}+M\delta k/k$ and likewise for $\delta J$.
It follows that
$ I_\infty = \delta M - \Omega\delta J - (M-\Omega J){\delta k\over k} +I^\prime_\infty$.
A computation using the asymptotic metric (\ref{falloff}) yields
\begin{equation}\label{iprime}
I_\infty^\prime = {A_{D-2}\over 16\pi}\left(  [(D-2)c_t+c_r] 
+ 4\Omega c_{t\phi} 
\right)\delta k + I_\infty^{div},
\end{equation}
where $I_\infty^{div}$ is a divergent term reflecting the infinite change in mass of the conical singularity on the azimuthal axis under the variation $\delta k$.  This term arises even in the absence of a black hole and will cancel against a similar divergence in $I_{axis}$.  It can be re-expressed as an axis integral 
$I_\infty^{div} = -{\delta k\over 4}\int dr\, dy^k\, r^{D-4}\,\sqrt{\bar\gamma}$.

The final task is to compute the boundary term $I_{axis}$ on the portion of azimuthal axis outside the black hole horizon. 
Using the near axis form of the metric given in (\ref{nearaxis}) one finds that the axis integral is given by
\begin{equation}\label{axis2}
I_{axis}={\delta k\over 4}\int_{axis} dr\, dy^k\, r^{D-4} \sqrt{-g_{tt}g_{rr}\gamma}\,
\end{equation}
The quantity $I_{axis}$ by itself is divergent,  but the combination $I_{axis}+I_\infty^{div}$ is easily seen to be finite.

Assembling our results above for $I_{h}$, $I_\infty$ and $I_{axis}$ gives the first law including variations in the angular range
\begin{equation}\label{firstlaw}
\delta M= {\kappa\delta A\over 8\pi} + \Omega\delta J +\left( 2\pi k\, \tphi + \Omega J\right) {\delta k\over k}
\end{equation}
where $2\pi k\,\tphi = 2\pi k\,\tphi^\infty + (I_{axis}+I_\infty^{div})/\delta k$.  The contribution to the angular tension from infinity matches that  found in the ADM construction (\ref{infinity}), while the axis contribution requires knowledge of the black hole metric in the interior of the spacetime.  The $\Omega J\delta k/k$ term in (\ref{firstlaw}) mirrors a contribution to the linear tension for boosted black strings \cite{Kastor:2007wr}.

\vskip 0.1in\noindent\textbf{Smarr relations:}  A simple formula for the angular tension, which makes computation of $I_{axis}$ in (\ref{axis2}) unnecessary, can be found from a combination of
Smarr relations.   The usual Smarr formula for stationary black holes $(D-3)M = (D-2)(\kappa A/8\pi+\Omega J)$
follows as the  condition for the first law to hold under an overall variation in length scale. 
Because the parameter $k$ is dimensionless, the new $\delta k$ terms in (\ref{firstlaw}) do not affect this result.

A second, independent relation  follows from a variation that is solely  $\delta k$.  Because the mass, horizon areas and angular momentum all depend linearly on $k$, one has $\delta M/M=\delta A/A=\delta J/J = \delta k/k$.  For the first law to hold under a variation $\delta k$, a stationary black hole black hole must therefore satisfy
\begin{equation}\label{newsmarr}
M = {\kappa A\over 8\pi} +2\Omega J +2\pi k\, \tphi 
\end{equation}
Using the first Smarr formula to eliminate the horizon area, one arrives at the general result for the angular tension
\begin{equation}\label{genten}
2\pi k\, \tphi = {M\over D-2} -\Omega J
\end{equation}
We have checked this formula via direct computation of $I_{axis}$ for Myers-Perry black holes.  We also find agreement with the static case (\ref{static}) when $J=0$.  Of course (\ref{newsmarr}) and (\ref{genten}) hold for $k=1$, the case of primary interest.

\begin{figure}[h]
\begin{center}\includegraphics[width=0.48\textwidth]{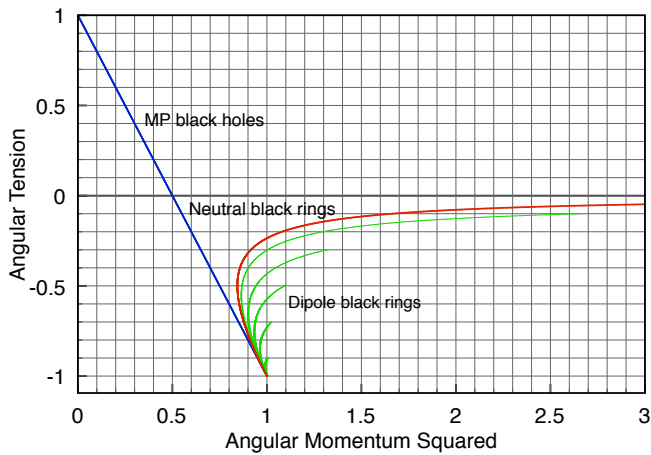}\end{center}
\caption{{\normalsize Phase diagram for singly spinning $D=5$ black holes with reduced angular tension $t_\phi$ plotted against the square of the reduced spin $j^2$ for $S^3$ black holes (blue), neutral black rings (red) and dipole black rings (green) for $\mu=0.1, 0.3,0.5, 0.7$ and $0.9$.}} 
\label{tensionfig}
\end{figure}

\vskip 0.1in\noindent\textbf{Phase diagram:}  Stationary black holes in $D=5$ are not uniquely characterized by $M$ and $J$.   For fixed $M$, spherical black holes coexist over a range of $J$ with two neutral black ring solutions \cite{Emparan:2001wn}.  Dipole black rings \cite{Emparan:2004wy} add a continuous degree of non-uniqueness, while black saturn \cite{Elvang:2007rd} and other multi-horizon spacetimes provide non-uniqueness over the whole range of angular momentum.  To what extent does adding angular tension as a new coordinate of the phase diagram distinguish between these different configurations?  In figure (\ref{tensionfig}) we plot results for spherical Myers-Perry black holes, neutral black rings, and dipole black rings.  Inclusion of multi-horizon spacetimes, especially those having unequal horizon angular velocities, will require further analysis.  
We define a dimensionless `reduced' angular tension variable $t_\phi$ and a reduced angular momentum $j$ according to
\begin{equation}\label{reduce}
t_\phi = 6\pi\tphi/M,\qquad j^2=27\pi J^2/32 M^3
\end{equation}
{\sl Spherical black holes:}   Singly spinning Myers-Perry black holes \cite{Myers:1986un} in $D=5$ have a finite range of angular momentum, $0\le j^2< 1$, with $j^2=1$ giving a naked singularity.   One finds that $t_\phi = 1-2j^2$ and therefore this branch is a straight line in figure (\ref{tensionfig}) which plots $t_\phi$  versus $j^2$.

\vskip 0.05in\noindent{\sl Neutral black ring:}   There is a one parameter family of balanced neutral black rings, having horizon topology $S^2\times S^1$ \cite{Emparan:2001wn}, indexed by $0\le \nu <1$.  
The reduced angular momentum  $j^2=(1+\nu)^3/8\nu$ and tension
$t_\phi = -\nu$, are plotted in red in figure (\ref{tensionfig}). 
Focusing on the range $27/32<j^2<1$, we see that the spherical black hole and black rings having the same angular momentum are distinguished by their angular tension. 

In the ultra-spinning regime $j\rightarrow\infty$ \cite{Emparan:2003sy}, the radius of the black ring becomes large and the ring is locally well-approximated by a boosted straight black string \cite{Hovdebo:2006jy} with vanishing linear tension \cite{Elvang:2003mj}.  Figure (\ref{tensionfig}) shows that the angular tension matches the linear tension in this ultraspinning limit.  For smaller black rings the angular tension becomes increasingly negative, corresponding to a positive outward pressure needed to balance the mutual gravitational attraction of ring segments.  
We have also checked that the angular tension (\ref{genten}) vanishes for the the approximate higher dimensional, ultra-spinning black ring solutions given in \cite{Emparan:2007wm}.

\vskip 0.05in\noindent{\sl Dipole black rings:} Black rings carrying a dipole charge $Q$ \cite{Emparan:2004wy} yield a continuous family of solutions having the same angular momentum over the range $27/32<j^2<\infty$.  
The dipole field contributes a $\Phi\delta Q$ term to the first law \cite{Copsey:2005se} leading to the modified Smarr formula $2M = 3({\kappa A\over 8\pi} + \Omega J) +\Phi Q$ \cite{Emparan:2004wy}
where $\Phi$ is a certain potential difference between the horizon and infinity.   The $3$-form field sourcing the dipole charges has not been considered above.  However, because $Q$ is obtained by integrating over an $S^2$ surrounding the ring and not over the $S^1$ whose range is being varied, we expect that the analysis would be unchanged.
Retracing the steps using Smarr relations then gives $2\pi\tphi = {1\over 3}(M+\Phi Q)-\Omega J$.

The family of balanced dipole rings 
includes a charge parameter with range $0\le\mu <1$, where $\mu=0$ gives the neutral black ring   
(the dilation coupling  has been set to $1$).  
The green lines in figure (\ref{tensionfig}) show results for a range of $\mu$ values, where the  
reduced tension (\ref{reduce}) has been generalized to $t_\phi= 6\pi\tphi/(M+\Phi Q)$.
Each line has a maximum value $j^2$ \cite{Emparan:2004wy}.
We see that introducing the angular tension indeed removes the infinitive degeneracy of the dipole rings, with dipole rings filling a region of the plane extending downward from the neutral black ring line.  
For fixed $j$, $t_\phi$ becomes increasingly negative as $\mu$ increases in order to balance the additional mutual attraction of  different portions of the ring from the $3$-form charge  \cite{Emparan:2004wy}.

\vskip 0.1in\noindent\textbf{Discussion:} Including angular tension as a new coordinate in the phase diagram for singly spinning $D=5$ black holes resolves the degeneracy between solutions having the same value of $j$.  We envision a number of possible stages of generalization for this analysis.  First, it will be interesting to include multiply spinning black holes, including angular tensions in each of the independent rotational planes.  Adding mixing (and variations in mixing) between the angular directions would further lead to an angular tension matrix in analogy with the linear tension matrix in Kaluza-Klein theory with multiple compact directions \cite{Kastor:2007wr}. It should also be possible to unify the linear and angular tensions in a single formalism for stationary Kaluza-Klein black holes, with mixing between the angular and compact directions corresponding to background Kaluza-Klein magnetic fields \cite{Dowker:1993bt,Dowker:1995gb}.

Unlike other ADM charges, angular tension depends on the metric in the interior of the spacetime at rotational axes.  The rod structure, or more generally the domain structure, of higher dimensional stationary black holes  \cite{Emparan:2001wk,Harmark:2004rm}, which provide  key elements in the classification of higher dimensional black holes \cite{Hollands:2007qf}, and are similarly based on data at such surfaces.  It will be interesting to see whether the rod/domain structure can be restated in the language of ADM charges by making use of  angular tension and its generalizations.

\vskip 0.1in\noindent{\bf Acknowledgments:}  We thank Jay Armas, Ted Jacobson and Niels Obers for helpful conversations and the organizers of the KITP workshop on \textit{Bits, Branes and Black Holes}, where this work was initiated, for hospitality.  This work was supported in part by NSF grant PHY-0555304.

\end{document}